\documentclass[aip, reprint, nofootinbib]{revtex4-1}
\usepackage{graphicx}

\usepackage[toc, section=section, nogroupskip, acronym]{glossaries}

\usepackage{amsmath}
\usepackage{amssymb,amsmath}
\usepackage[usenames,dvipsnames]{color}
\usepackage{cancel}
\usepackage[margin=.7in]{geometry}
\usepackage{braket}
\usepackage{enumitem}
\usepackage{mathtools}
\usepackage{setspace}
\usepackage[english]{babel} 
\usepackage[titletoc]{appendix}
\usepackage{titlesec}
\usepackage{hyperref}
\usepackage{ntheorem}
\theoremseparator{:}
\usepackage{ulem}

\usepackage{acro}
\usepackage[nottoc]{tocbibind}
\usepackage[many]{tcolorbox}
\usetikzlibrary{decorations.pathreplacing}

\usepackage{filecontents}
\pdfoutput=1

\newtcolorbox{leftbrace}{%
    enhanced jigsaw, 
    breakable, 
    frame hidden, 
    overlay={%
        \draw [
            decoration={brace,amplitude=0.5em},
            decorate,
            ultra thick,
        ]
        (frame.north west)--(frame.south west);
    },
    parbox=false,
}

\begin{document}
\normalem
\title{On the Formation of Solid States Beyond Perfect Crystals: Quasicrystals, Geometrically-Frustrated Crystals and Glasses}

\author{Caroline S. Gorham}
\email{caroling@cmu.edu}
\affiliation{Department of Materials Science and Engineering, Carnegie Mellon University, Pittsburgh, PA 15213, USA}

\author{David E. Laughlin}
\email{laughlin@cmu.edu}
\affiliation{Department of Materials Science and Engineering, Carnegie Mellon University, Pittsburgh, PA 15213, USA}

\begin{abstract}
There are three kinds of solid states of matter that can exist in physical space: quasicrystalline (quasiperiodic), crystalline (periodic) and amorphous (aperiodic). Herein, we consider the degree of orientational order that develops upon the formation of a solid state to be characterized by the application of quaternion numbers.  The formation of icosahedral quasicrystalline solids is considered alongside the development of bulk superfluidity, characterized by a complex order parameter, that occurs by spontaneous symmetry breaking in three-dimensions. Crystalline solid states are viewed as higher-dimensional analogues to phase-coherent topologically-ordered superfluid states of matter that develop in restricted dimensions (Hohenberg-Mermin-Wagner theorem). Lastly, amorphous solid states are viewed as dual to crystalline solids, in analogy to Mott-insulating states of matter that are dual to topologically-ordered superfluids.
\end{abstract}
\maketitle

\section{Introduction}

The sphere-packing problem, which describes the arrangement of atomic constituents in close-packed crystalline solid states, has been of great interest for several centuries~\cite{knudson_stacking_2019}. In three-dimensions, this topic came to be known as Kepler's conjecture~\cite{kepler_strena_1611, kepler_six-cornered_2010} which states that no arrangement of equally-sized spheres filling space in three-dimensions has a greater average density~\cite{gauss_Untersuchungen_1831} than $\pi/3\sqrt{2}\approx 0.74$.  Cubic close-packing (FCC) and hexagonal close-packing (HCP) arrangements are examples of such sphere-packing in three-dimensions. This conjecture was proved computationally by Hales in 1998~\cite{hales_kepler_1998, sloane_keplers_1998, hales_proof_2005}. More recently, Viazovska~\cite{viazovska_sphere_2017, cohn_conceptual_2017} proved that the closest-packing of equally-sized spheres in eight-dimensions is the $E8$ lattice which has a density of $\pi^4/384\approx 0.25$. These close-packed arrangements, of equally-sized spheres, describe crystalline ground states with perfect orientational order in either three-dimensions or eight-dimensions.

It has been proposed by the authors that the problem of solidification can be understood by the application of a \emph{quaternion} ($\mathbb{Q}$) order parameter, to characterize the degree of orientational order in the solid states that form from undercooled liquids.  In this way, it is possible to frame crystalline solids (orientationally-ordered) in three-dimensions or eight-dimensions as generalizations of phase-coherent superfluids~\cite{kosterlitz_ordering_1973} that are characterized by a \emph{complex} ($\mathbb{C}$) order parameter~\cite{london_superfluids_1961}. Following this approach, we have generated a unified topological framework within which to understand the formation of the three types of solid states in three-dimensions:
\begin{itemize} 
\item quasicrystalline~\cite{gorham_solidification_2019} (quasiperiodic)
\item crystalline~\cite{gorham_topological_2019, gorham_crystallization_2019} (periodic) 
\item amorphous~\cite{gorham_topological_2019} (aperiodic)
\end{itemize}
This unified topological framework, that applies to the solidification of the three kinds of solid states of matter (listed above) is constructed via the application of \emph{topology} and the theory of \emph{Universality classes of phase transitions}~\cite{herbut_modern_2007}. 

Perfectly orientationally-ordered crystalline structures, i.e., those without any topological defects, are idealized cases of solids that can only be realized at 0 K. At temperature, even these perfect crystalline solids have a finite density of topological defects (e.g., dislocations) whose nucleation is necessary in order to account for crystal entropy. Beyond perfectly orientationally-ordered crystals, \emph{geometrically-frustrated} crystalline solids~\cite{frank_complex_1958, frank_complex_1959, sadoc_geometrical_2006} (i.e., topologically close-packed, TCP) that possess 5-fold local icosahedral order~\cite{frank_supercooling_1952} contain topological defects in the ground state and warrant considerable attention. These geometrically-frustrated crystalline solids, which have a periodic spatial distribution of orientational disorder in the ground state, can also be understood within our unified topological framework for solidification. Thus, by considering a quaternion orientational order parameter we go beyond the conventional sphere-packing problem that applies to perfect crystalline solids in order to consider the formation of solid states with a spatial distribution of orientational disorder (curvature).

The main section of this article (Section~\ref{sec:solidification}) focuses on a topological ordering field theory for solidification, using quaternion numbers to characterize orientational order. Section~\ref{sec:solidification} presents a discussion on the mechanisms of formation of the three types of solid states of matter. Section~\ref{sec:solidification}  is separated into two distinct subsections that describe solidification in ``bulk" dimensions (Section~\ref{sec:quasicrystals}) and in ``restricted" dimensions (Section~\ref{sec:crystal-to-glass}) separately. Section~\ref{sec:quasicrystals} describes the solidification of icosahedral quasicrystalline states of matter, which derive by projection from the eight-dimensional E8 lattice, in analogue to the formation of bulk superfluids in three-dimensions. Solidification of crystalline and non-crystalline solids, in the vicinity of an anticipated crystalline-to-glass quantum phase transition that may be realized in ``restricted" dimensions, are discussed in Section~\ref{sec:crystal-to-glass}.

\section{Quaternion Field Theory of Solidification}
\label{sec:solidification}

\subsection{Solidification in ``Bulk'' Dimensions: Quasicrystals}
\label{sec:quasicrystals}

Quaternion numbers (SU(2)) have a 2-to-1 homomorphism~\cite{mermin_homotopy_1978, mermin_topological_1979} with the group of rotations about an origin in three-dimensional Euclidean space ($SO(3)$), and may therefore be used to characterize orientational order that develops upon \emph{solidification}. Quaternion numbers are four-dimensional, such that the group of all unit quaternions is the hyperspherical  surface ($S^3\in\mathbb{R}^4$) of constant positive curvature. The discrete nature of orientational order in crystalline solid states is accounted for by considering the ground state manifold to be $\mathcal{M}=G/H'$, where G=SU(2) and $H'$ is the \emph{binary polyhedral group} representation of the preferred orientational order group $H\in SO(3)$. The relevant ground state manifold $\mathcal{M}$ rests in the quaternion plane. Thus, the main kind of available topological defects belong to the \emph{third homotopy group} (i.e., $\pi_3(\mathcal{M})$). 

Third homotopy group topological defects are higher-dimensional analogues to quantum vortices in superfluid ordered states of matter, in which a complex order parameter characterizes the degree of order (`Mexican hat' free energy function~\cite{endres_higgs_2012}). Quantum vortices in superfluids are holes with the superfluid circulating around the vortex axis~\cite{annett_superconductivity_2004}. Just as the free energy cost to introduce a vortex line in three-dimensional superfluids is much too high for them to appear in the absence of external fields~\cite{halperin_resistive_1979}, the same argument applies for third homotopy group topological defects in dimensions larger than four. On the other hand, third homotopy group defects are points in the four-dimensional quaternion plane and so, the free energy cost to introduce these defects in four-dimensions should be comparable to $\text{k}_\text{B}\text{T}$ in the absence of external fields~\cite{halperin_resistive_1979}. 

It has long been understood that icosahedral quasicrystalline (quasiperiodic) solids can be derived from the 8-dimensional lattice E8, whose close-packed neighbor shells are embedded in 7-dimensional spheres ($S^7$). This is described mathematically by making use of the \emph{quaternion Hopf fibration}~\cite{sadoc_e8_1993} decomposition of $E8$. Third homotopy group topological defects in eight-dimensions are higher-dimensional than points, and it is therefore anticipated -- in analogy to vortex lines in superfluids -- that they may only be introduced in the presence of an applied field~\cite{halperin_resistive_1979}. In this way, in eight-dimensions, the presence of third homotopy group defects should not prevent \emph{spontaneous symmetry breaking} at the melting temperature. 

Thereby, it is anticipated that, the E8 lattice may be realized by a conventional disorder-order phase transition that occurs in the ``bulk" dimension for the quaternion orientational order parameter~\cite{gorham_solidification_2019}. Quasiperiodic structures in physical space are therefore considered to be most similar to superfluid ordered systems that exist in three-dimensions~\cite{gorham_solidification_2019}, which do not form by topological-ordering (in the Berezinskii-Kosterlitz-Thouless sense~\cite{berezinskii_destruction_1971, kosterlitz_ordering_1973}).

\subsection{Solidification in ``Restricted" Dimensions 4D/(3D+1t): Crystal-to-Glass Quantum Phase Transition}
\label{sec:crystal-to-glass}

Phase transitions, and transport properties of ordered states of matter become more interesting when the degrees of freedom of the system of particles that undergo ordering are ``restricted" in a dimensional sense (\emph{Hohenberg-Mermin-Wagner} (HMW) \emph{theorem}~\cite{mermin_absence_1966, hohenberg_existence_1967, halperin_hohenbergmerminwagner_2018}). The original HMW theorem states that any system with continuous symmetry ($O(N)$, $N\geq 2$) cannot undergo \emph{spontaneous symmetry breaking} (SSB) at any finite temperature in two- or one-dimensions~\cite{mermin_absence_1966}. In this way, 2D/(1D+1t) may be considered to be ``restricted'' (because SSB is not possible) for any ordered system that breaks a continuous symmetry group. As a consequence of the original HMW theorem, traditional phase-coherent ordered states cannot occur by any conventional disorder-order phase transition in two- or one-dimensions~\cite{lee_symmetry_1986}.

However, low-temperature complex ordered superfluids do exist in ``restricted" dimensions. In the case of complex ordered systems (i.e., $N=2$) that exist in 2D/(1D+1t), which belong to the ($2, 2$) Universality class, the nature of the novel phase transition towards the phase-coherent superfluid ground state was explained in terms of the topological-ordering of vortex/anti-vortex point defect pairs by Berezinskii~\cite{berezinskii_destruction_1971} and Kosterlitz and Thouless~\cite{kosterlitz_ordering_1973}. Such complex ordered systems, that exist in ``restricted" dimensions, are described mathematically by the application of the $O(2)$ quantum rotor model~\cite{vojta_quantum_2006, sachdev_quantum_2011, endres_higgs_2012}. 

Such $O(N)$ quantum rotor models are essential for the study of quantum phase transitions~\cite{sachdev_quantum_2011}, between phase-coherent and phase-incoherent low-temperature ordered states. In particular, in the two-dimensional $N=2$ case, the $O(2)$ quantum rotor model predicts the existence of a quantum phase transition between the superfluid and a Mott-insulator ordered state~\cite{endres_higgs_2012} that has been realized experimentally. In charged superfluids, this is the superconductor-to-superinsulator transition~\cite{vinokur_superinsulator_2008, baturina_superinsulatorsuperconductor_2013, sankar_disordered_2018, diamantini_superconductor-superinsulator_2018}.

In the following subsections (\ref{sec:xtal}, \ref{sec:fk}, \ref{sec:glass}), we summarize our recent generalizations of the notion of ``restricted" dimensions~\cite{gorham_su2_2018, gorham_topological_2019, gorham_crystallization_2019} (in the HMW theorem sense) to systems with continuous symmetry ($O(N)$, $N\geq 4$) that exist in 4D/(3D+1t). In quaternion ordered systems ($N=4$), this is a consequence of the nature of \emph{third homotopy group} topological defects as points in 4D/(3D+1t) which are spontaneously generated at finite temperatures. These topological defects generate an abundance of misorientational fluctuations in the quaternion order parameter, that prevent \emph{spontaneous symmetry breaking} at finite temperatures. This generalization of the original HMW theorem, to quaternion ordered systems, has been applied (by the authors) to understand the mechanisms of solidification of crystalline and non-crystalline solid states in the vicinity of a first-order quantum phase transition at the Kauzmann point~\cite{kauzmann_nature_1948}.

\subsubsection{Perfect crystalline solid states}
\label{sec:xtal}

Crystallization in 4D/(3D+1t) has been described as a defect-driven phase transition~\cite{gorham_su2_2018, gorham_topological_2019, gorham_crystallization_2019}, that occurs at a finite temperature below the melting temperature, in analogue to the formation of phase-coherent superfluidity in 2D/(1D+1t) systems. Therefore, crystallization in 4D/(3D+1t) belongs to the Berezinskii-Kosterlitz-Thouless universality class~\cite{berezinskii_destruction_1971, kosterlitz_ordering_1973} of phase transitions for quaternion ordered systems. This particular topological phase transition, occurs as third homotopy group point defects and anti-point defects bind into complementary pairs. This is a direct higher-dimensional analogue to vortex/anti-vortex binding at the prototypical BKT-transition in complex ordered thin-films~\cite{kosterlitz_critical_1974}.   

In addition to third homotopy group topological defects, as a consequence of discrete orientational order of atomic clusters in undercooled fluids~\cite{frank_supercooling_1952}, misorientational fluctuations that develop below the melting temperature also consist of disclinations that belong to the fundamental homotopy group of $\mathcal{M}$. Upon crystallization, the topology of $\mathcal{M}$ changes discontinuously as a result of the development of a translational lattice that corresponds to a Brillouin zone in reciprocal space. This change in the topology of $\mathcal{M}$, from a spherical to a toroidal manifold, constitutes a first-order phase transition and forces the formation of bound states of complementary disclinations that can be regarded as a dislocations~\cite{seung_defects_1988, chaikin_principles_2000, landau_theory_1986, gopalakrishnan_disclination_2013, pretko_fracton-elasticity_2018} that can be drawn around $\mathcal{M}$.

In this way, due to the formation of bound states of complementary curvature-carrying disclinations, crystallization may be viewed as a \emph{flattening} of an undercooled liquid into a solid state. Perfect orientational-order is obtained for crystalline ground states in the absence of geometrical frustration~\cite{sadoc_geometrical_2006}. This occurs because, in the absence of geometrical frustration, the plasma of topological defects that forms just below the melting temperature is perfectly balanced and all of the defects are able to form bound pairs. This ensures that the perfect crystalline ground state is free of permanent topological defects, and is therefore flat everywhere. Such bound states of disclinations, i.e., dislocations, are imperfections in a crystalline lattice and only become excited at finite temperatures. In the same way, bound states of third homotopy group topological defects are only present at temperatures above 0 K.

\subsubsection{Geometrically-frustrated crystalline solid states (TCP)}
\label{sec:fk}

In analogue to complex ordered systems that exist in ``restricted" dimensions, i.e., 2D/(1D+1t), \emph{frustration} of ground states becomes possible for quaternion ordered systems that exist in 4D/(3D+1t). Increasing frustration drives the ordered system towards a \emph{quantum phase transition}, that belongs to the relevant $O(N)$ quantum rotor model~\cite{sachdev_quantum_2011}. In the case of crystalline solid states, the relevant frustration parameter is geometrical~\cite{sadoc_geometrical_2006} and is proportional to the curvature that is associated with atomic vertices whose local orientational order is incompatible with creating a space-filling arrangement~\cite{nelson_symmetry_1984}. 

The most well-known manifestation of geometrical frustration in crystals occurs in Frank-Kasper phases of complex transition metal alloys~\cite{frank_complex_1958, frank_complex_1959}, that form from icosahedrally-coordinated undercooled fluids~\cite{nelson_symmetry_1984}. Positive curvature that is attributed to each icosahedrally-coordinated atomic vertex is inversely proportional to the radius of the $\{3,3,5\}$ polytope, which consists of 120 particles (the binary icosahedral group) inscribed on the surface of a sphere in four-dimensions. The finite positive curvature, associated with icosahedral clustering in undercooled fluids, biases the plasma of topological defects that comprise the gas of misorientational fluctuations below the melting temperature. This ensures that the space remains flat overall. 

Upon crystallization, these excess negative-signed topological defects cannot form bound states and must persist to the ground state. Topological defects in the crystalline ground state, induced by geometrical frustration, form a lattice in analogue to the Abrikosov flux lattice in topologically-ordered superconductors in the presence of an applied magnetic field~\cite{nelson_defects_2002}. In three-dimensional Frank-Kasper crystalline solids, this manifestation of geometrical frustration in the ground state is evidenced as the \emph{major skeleton network}~\cite{nelson_liquids_1983} of disclination lines that concentrate negative curvature and whose periodic arrangement satisfies the third law of thermodynamics. With a sufficient amount of geometrical frustration, the orientationally-ordered crystalline solid state becomes entirely broken down as the distance between topological defects becomes reduced below a critical value.

\subsubsection{Non-crystalline solid states}
\label{sec:glass}

Within this unified topological framework for solidification, amorphous solids are considered to be topologically-ordered phases that may arise in 4D/(3D+1t) uncharged quaternion ordered systems~\cite{gorham_topological_2019} in the vicinity of a crystalline-to-glass quantum phase transition that belongs to the $O(4)$ quantum rotor model. The possibility of the existence of non-crystalline solid states (i.e., orientationally-disordered) is due to the duality of the phase-amplitude uncertainty principle, that applies to the quaternion orientational order parameter wave-function. In this way, glassy solids are considered to be a higher-dimensional analogue to the Mott-insulator or superinsulator ground states~\cite{baturina_superinsulatorsuperconductor_2013} that emerge on the insulating side of the superfluid-to-Mott insulator or superconductor-to-superinsulator transition (SIT) in complex ordered thin films~\cite{diamantini_gauge_1996, vinokur_superinsulator_2008}.

A solidification phase diagram, in the vicinity of the first-order crystalline-to-glass quantum phase transition at the Kauzmann point, has been proposed by the authors~\cite{gorham_topological_2019, gorham_crystallization_2019}. The glass transition is considered to belong within the Berezinskii-Kosterlitz-Thouless (BKT) (4,4) Universality class, that applies to quaternion ordered systems that exist in ``restricted" dimensions. In this sense, the glass transition occurs via the condensation of topological defects and localization of atomic particles~\cite{gorham_topological_2019}. The glass transition in three-dimensions is analogous to the formation of superinsulating phases in Josephson junction chains, that occur via instanton condensation~\cite{polyakov_gauge_1987} and localization of Cooper pairs~\cite{diamantini_confinement_2018}.

The duality between ordered states of matter that can be realized in ``restricted" dimensions, in the vicinity of a quantum phase transition that belongs to the relevant $O(N)$ quantum rotor model~\cite{sachdev_quantum_2011, endres_higgs_2012}, manifests observably in the transport properties. For example, as originally predicted by t'Hooft~\cite{hooft_phase_1978} (1978), the superinsulator that is the dual to a superconductor exhibits infinite electrical resistance~\cite{diamantini_electrostatics_2019}. In the same way, the thermal conductivity of crystalline and glassy solid states have long been known~\cite{eucken_uber_1911, kittel_interpretation_1949} to exhibit inverse behavior above approximately 50 K. Many theoretical models exist to interpret the thermal conductivity of glasses in this temperature range, which decreases as a function of decreasing temperatures~\cite{einstein_elementary_1911,birch_thermal_1940,  kittel_interpretation_1949, slack_thermal_1979, cahill_thermal_1987, nakayama_increase_1999} -- above a plateau at low-temperatures that is a universal feature of the thermal conductivity in non-crystalline solids~\cite{zeller_thermal_1971, freeman_thermal_1986, graebner_phonon_1986,  leggett_amorphous_1991, ramos_low-temperature_1997}. The topological framework for solidification of crystals and glasses elucidated herein points towards a novel interpretation of this inverse transport behavior~\cite{gorham_topological_2019}.

\section{Conclusions}
\label{sec:summary}

In this treatise on solidification, a unified framework has been presented to consider the phase transitions that give rise to the three types of solid states of matter: quasicrystalline (Section~\ref{sec:quasicrystals}), crystalline and glassy (Section~\ref{sec:crystal-to-glass}). This approach to solidification relies on the application of a quaternion order parameter, to characterize the degree of orientational order that develops below the melting temperature. Fundamentally, orientationally-ordered crystalline solids are viewed as generalizations of phase-coherent superfluids. 

The Kauzmann point, that occurs at a finite temperature, has been associated with a quantum critical point that identifies a first-order quantum phase transition belonging to the $O(4)$ quantum rotor model. Such a finite-temperature quantum critical point, between crystalline and non-crystalline solid states, is classified as first-order. Topologically, this is owing to the fact that the genus of the orientational order parameter manifold changes discontinuously from a torus (in the crystal) to a sphere (in the glass). This quantum phase transition is a higher-dimensional analogue to the superfluid-to-Mott insulator transition. It follows that, solid states in the vicinity of the Kauzmann point exhibit topological-order in the Berezinskii-Kosterlitz-Thouless sense.

\section{Acknowledgements}
The authors gratefully acknowledge support from the ALCOA Chair in Physical Metallurgy. C.S.G. also thanks the lecturers and participants of Beg Rohu's Summer School 2019: Glasses, Jamming and Slow Dynamics, for insightful discussions regarding the glass transition.


\bibliography{\jobname}
\end{document}